\documentstyle[epsf,prl,aps]{revtex}   
\tightenlines   
\begin{document}   
\twocolumn[\hsize\textwidth\columnwidth\hsize\csname@twocolumnfalse\endcsname 
  
\preprint{\vbox{\hbox{November 1997}   
\hbox{IFP-746-UNC}  }}   
\title{Charged fermion masses with Yukawa coupling strength universality}
\author{M. D. Tonasse}  
\address{Instituto Tecnol\'ogico de Aeron\'autica, Centro T\'ecnico  
Aeroespacial, Pra\c ca Marechal do Ar Eduardo Gomes 50, 12228-901 S\~ao Jos\'e 
dos Campos, SP, Brazil}  
\date{\today}  
\maketitle    
\begin{abstract}  
We consider the problem of the ordinary charged fermion mass patterns in the framework of a version of the 3-3-1 electroweak gauge model, which includes charged heavy leptons. The masses of the top and the bottom quarks are given at the tree level. All the other charged fermions get their masses at the one loop level. The Yukawa coupling strength can be universal. The model is compatible with a mechanism for neutrino mass generation proposed by other authors.
\end{abstract}
\vskip 1.1 pc]



\narrowtext

\section{Introduction}

The pattern of the fermion masses and mixing angles is one of the major problems of the standard electroweak model. The standard model is based on SU(3)$_C\otimes$SU(2)$_L\otimes$U(1)$_Y$ symmetry group and contains only one Higgs scalar doublet of which only one neutral scalar field remains after the symmetry breakdown. In this model, the fermion masses are given by $m_f = g_f\,v_W/\sqrt{2}$, where $g_f$ is the Yukawa coupling constant associated with the fermion $f$ and $v_W$ is the vacuum expectation value (VEV) of the Higgs field in the theory. Therefore, the correct values of the fermion masses in the standard model require appropriate choice of the Yukawa parameters, making $g_f$ to run in a range of ${\cal O}\left(10^{-6}\right)$ (for the electron) to ${\cal O}\left(1\right)$ (for the top quark). All the fermion masses, except to the quark top one, lie very below of the dynamic scale $v_W = 246$ GeV. Such problems and other undesired features support the belief that the standard model is only an effective low energy electroweak theory and give motivation for developing extensions and alternative models.\par
There are essentially two attempts to understand the fermion mass and mixing patterns. One of them consists in searching the answer in more fundamental theories such as grand unified theories \cite{AC93}, supersymmetric left-right \cite{BR77} and composite models \cite{BP91}, etc. The alternative road is to construct phenomenological mass matrices, by employing some specific {\it ansatz}, attempting to fit low energy data \cite{IO98}.\par
In the present paper, we address the problem of the charged fermion masses in the context of the SU(3)$_C\otimes$\-SU(3)$_L\otimes$\-U(1)$_N$ (3-3-1 for short) model in which the weak and electromagnetic interactions are described through a gauge theory based on the SU(3)$_L\otimes$U(1)$_N$ semi simple symmetry group \cite{FH93}. The most interesting feature of the model is that the anomaly cancellations occur only when the three fermion generations are considered together and not family by family as in the standard model. This implies that the number of families must be a multiple of the color number and, consequently, the 3-3-1 model suggests a route towards the response of the flavor question \cite{FH93}. The model has also a great phenomenological interest since the related new physics can occur in a scale near of the Fermi one. Therefore, since in this model individual lepton number conservation can be violated, typical 3-3-1 processes, which are free of the standard model background, can be studied in the next generation of colliders \cite{CQ99}.\par
There are several versions of the 3-3-1 model and all of them preserve the essential features discussed above \cite{FH93,MP93,PT93}. In this work we are interested in a version which includes heavy leptons \cite{PT93}. It is interesting to notice that these heavy leptons do not belong to any of the four types of heavy leptons usually considered in the literature, {\it i. e.}, (i) sequential leptons, (ii) paraleptons, (iii) ortholeptons, or (iv) long-lived penetrating particles. Hence, the experimental limits already existing do not apply directly to them \cite{PT93,Gea00}.\par
Fermion masses and the Cabibbo-Kobayashi-Maskawa mixing matrix problem have already been considered in another version of the 3-3-1 model in Ref. \cite{TJ98} where an SU(2) horizontal symmetry was imposed and all masses of the first family vanish.\par

\section{Basic facts about 3-3-1 heavy lepton model}

Let us summarize the most relevant points of the original 3-3-1 heavy lepton model. The left-handed lepton and quark fields form SU(3)$_L$ triplets, {\it i. e.}, 
\begin{mathletters}
\begin{eqnarray}
\psi_{aL} = \pmatrix{\nu^\prime_a \cr l^\prime_a \cr P^\prime_a \cr}_L, & \quad & Q_{1L} = \pmatrix{u^\prime_1 \cr d^\prime_1 \cr J_1 \cr}_L, \\
Q_{\alpha L} & = & \pmatrix{J^\prime_\alpha \cr u^\prime_\alpha \cr d^\prime_\alpha \cr}_L, 
\label{ferm}\end{eqnarray}\end{mathletters}\noindent
which transform as $\left({\bf 3}, 0\right)$, $\left({\bf 3}, 2/3\right)$ and $\left({\bf 3}^*, -1/3\right)$, respectively, where $l_a = e, \mu , \tau$, and 0, 2/3 and $-$1/3 are the U(1)$_N$ charges\cite{PT93}. Throughout this work we are using the convention that the Latin indices run over 1, 2, and 3, while the Greek ones are 2 and 3. Each charged left-handed fermion field has its right-handed counterpart transforming as a singlet of the SU(3)$_L$ group. Right-handed neutrinos are optional in the model. In order to avoid anomalies one of the quark families, no matter which, must transform in a different way with respect to the other two. The $J_1$ exotic quark carries 5/3 units of elementary electric charge, while $J_2$ and $J_3$ carry $-$4/3 each. The primed fields are the interaction eigenstates, which are related with the respective mass eigenstates by
\begin{mathletters}
\begin{eqnarray}
\nu_{aL} = {\tt N}_{ab}\nu_{bL}, & \qquad & l_{aL, R}^\prime = {\tt L}^{L, R}_{ab}l_{bL, R}, \\
P^\prime_{aL, R} = {\tt P}^{L, R}_{ab}P_{bL, R}, & \qquad &
U^\prime_{aL, R} = {\tt U}^{L, R}_{ab}U_{bL, R}, 
\label{mix1}\\ 
D^\prime_{aL, R} = {\tt D}^{L, R}_{ab}D_{bL, R}, & \qquad & J^\prime_{\alpha L, R} = {\tt J}^{L, R}_{\alpha\beta}J_{\beta L, R},
\label{mix2}\end{eqnarray}\end{mathletters}\noindent
where ${\tt N}$, ${\tt L}^{L, R}$, ${\tt P}^{L, R}$, ${\tt U}^{L, R}$, ${\tt D}^{L, R}$ and ${\tt J}^{L, R}$ are unitary mixing matrix. We define the physical quark fields as $U_{L, R} = \pmatrix{u & c & t}_{L, R}^T$, $D_{L, R} = \pmatrix{d & s & b}_{L, R}^T$ and $J_{L, R} = \pmatrix{J_1 & J_2 & J_3}_{L, R}^T$ with their analogues for the interaction eigenstates. \par
In the gauge sector the model predicts, in addition to the standard $W^\pm$ and $Z^0$, the extras $V^\pm$, $U^{\pm\pm}$ and the ${Z^\prime}^0$ gauge bosons.\par

The fermion and gauge boson masses are generated in the model through the three Higgs scalar triplets 
\begin{equation}
\eta = \pmatrix{\eta^0 \cr \eta_1^- \cr \eta^+_2 \cr}, \quad \rho = \pmatrix{ \rho^+ \cr \rho^0 \cr \rho^{++} \cr}, \quad \chi = \pmatrix{ \chi^- \cr \chi^{--} \cr \chi^0 \cr},
\label{tri}\end{equation}
which transform under the SU(3) group as $\left({\bf 3}, 0\right)$, $\left({\bf 3}, 1\right)$ and $\left({\bf 3}, -1\right)$, respectively. The neutral scalar fields develop the VEVs $\langle\eta^0\rangle = v$, $\langle\rho^0\rangle = u$, and $\langle\chi^0\rangle = w$, with
\begin{equation}
v^2 + u^2 = v_W^2.
\label{vev}\end{equation}
The pattern of symmetry breaking is SU(3)\-$_L\otimes$\-U(1)$_N$\-$\stackrel{\langle\chi\rangle}{\longmapsto}\- $SU(2)\-$_L\otimes$\-U(1)$_Y$\-$\stackrel{\langle\rho, \eta\rangle}{\longmapsto}$\-U(1)$_{{\rm em}}$ and so we can expect $w >> v, u$. In the original heavy lepton model from Ref. \cite{PT93} the $\eta$ and $\rho$ give masses to the ordinary quarks and the charged lepton masses come from the $\rho$ Higgs multiplet. The masses of all the exotic fermions rise through the $\chi$ triplet. The neutrinos can gain their masses at the tree level by the $\eta$ Higgs triplet. In the original 3-3-1 heavy lepton model the full Yukawa Lagrangians, which must be considered are
\begin{mathletters}
\begin{eqnarray}
{\cal L}_l & = & -\sum_{ab}\left(\frac{1}{2}\epsilon^{ijk}G^{\left(\nu\right)}_{ab}\overline{{\psi_{aiL}}^C}\psi_{bjL}\eta_k + \right. \cr
&& \left. G^{\left(l\right)}_{ab}\overline{\psi}_{aL}l^{\prime-}_{bR}\rho - G^{\left(P\right)}_{ab}\overline{\psi}_{aL}P^{\prime+}_{bR}\chi\right)
\label{yukl} + {\mbox{H. c.}}, \\
{\cal L}_Q & = & \overline{Q}_{1L}\sum_b\left(G^{\left(U\right)}_{1b}U^\prime_{bR}\eta + G^{\left(D\right)}_{1b}D^\prime_{bR}\rho + \right. \cr
&& \left. G^{\left(J\right)}J_{1R}\chi\right) + \sum_\alpha\overline{Q}_{\alpha L}\left(F^{\left(U\right)}_{\alpha b}U^\prime_{bR}\rho^* + \right. \cr
&& \left. F^{\left(D\right)}_{\alpha b}D^\prime_{bR}\eta^* + \sum_{\beta}F^{\left(J\right)}_{\alpha\beta}J^\prime_{\beta R}\chi^*\right) + \mbox{H. c.},
\label{yukq}\end{eqnarray}\end{mathletters}\noindent
where $G^{\left(\nu\right)}_{ab}$, $G^{\left(l\right)}_{ab}, G^{\left(P\right)}_{ba}$, $G^{\left(U\right)}_{1b}$, $F^{\left(U\right)}_{\alpha b}$, $G^{\left(D\right)}_{1b}$, $F^{\left(D\right)}_{\alpha b}$, $G^{\left(J\right)}$ and $F^{\left(J\right)}_{\alpha\beta}$ are the Yukawa coupling constants. SU(3)$_C$ indices have been suppressed and $\eta^*$, $\rho^*$ and $\chi^*$ denote the $\eta$, $\rho$ and $\chi$ antiparticle fields, respectively. \par
The physical massive eigenstates $H^+_1$, $H^+_2$ and $H^{++}$ of the charged Higgs bosons of the model are defined by
\begin{mathletters}
\begin{eqnarray}
\pmatrix{\eta_1^+ \cr \rho_+ \cr} & = & \frac{1}{v_W}\pmatrix{-v & u \cr u & v\cr}\pmatrix{G^+_1 \cr H^-_1 \cr},\\
\pmatrix{\eta^+_2 \cr \chi^+ \cr} & = & \frac{1}{\sqrt{v^2 + w^2}}\pmatrix{-v & w \cr w & v\cr}\pmatrix{G^+_2 \cr H^-_2 \cr}, \\
\pmatrix{\rho^{++} \cr \chi^{++} \cr} & = & \frac{1}{\sqrt{u^2 + w^2}}\pmatrix{-u & w \cr w & u\cr}\pmatrix{G^{++} \cr H^{++} \cr},
\end{eqnarray}\label{higgs}\end{mathletters}\noindent
where $G_1^+$, $G_2^+$ and $G^{++}$ are the massless charged Goldstone bosons\cite{TO96}. \par
We notice that from Eq. (\ref{yukq}) quarks of the same charge are coupled through different Higgs multiplets in Eqs. (\ref{tri}). This leads to the undesired flavor changing neutral currents (FCNCs) at the tree level with the couplings of the quarks through the Higgs and the extra neutral gauge boson ${Z^{\prime }}^{0}$, but not through those to the standard $Z^{0}$. They appear also in the Yukawa interactions involving exotic quarks. This problem was extensively studied in several papers where is shown that the Glashow-Iliopoulos-Maiani mechanism can be implemented and the FCNCs are naturally suppressed in the model \cite{MP93,OZ96}.\par

\section{Charged fermion masses}
\label{secIII}

Let us now study the consequences of the model to the quark and lepton masses. 
In this paper we concentrate in the Yukawa sector of the model and we implement a scheme to generating charged ordinary mass terms with Yukawa universality. Thus, in order to describe the mass patterns we see from Eqs. (\ref{yukq}) that some of them must be generated at loop level. It is possible since we allow all the exotic fermions to gain their masses at tree level {\it via} the $\chi$ Higgs triplet.\par 
We analyze firstly the quark sector. We note that if the $S_3$ permutation symmetry operates within each quark family in Eq. (\ref{yukq}) before the symmetry breaking we have not Yukawa hierarchy in each charge sector of the quarks at the tree level. In addition, if we impose also $S_{3L}\otimes S_{3R}$ permutation symmetry among the quark families and the $S_3$ symmetry among the Higgs multiplets, supplemented by the set of discrete symmetry 
\begin{mathletters}
\begin{eqnarray}
\eta \to \rho^*, \quad \rho \to \eta^*, & \quad & \chi \to \chi^*, \quad Q_{aL} \to Q_{aL}, \\
U^\prime_R \to U^\prime_R, & \quad & D^\prime_R \to D^\prime_R, \quad J^\prime_R \to J^\prime_R, 
\end{eqnarray}\label{simq}\end{mathletters}\noindent
the Yukawa coupling strength universality is implemented in all quark sector. Therefore, the mass matrices for the 2/3 and the $-$1/3 charge sector can be written as
\begin{mathletters}
\begin{eqnarray}
\Gamma_{U} & = & G\pmatrix{v & v & v \cr u & u & u \cr u & u & u \cr}, 
\label{mata}\\
\Gamma_{D}& = & G\pmatrix{-u & -u & -u \cr v & v & v \cr v & v & v \cr}, \label{matb}
\end{eqnarray}\label{mat}\end{mathletters}\noindent
respectively. Thus, only two quarks, which we identify as the top and the bottom one, gain masses at the tree level. Therefore, from Eqs. (\ref{mat}) we have
\begin{equation}
v = \frac{2m_t - m_b}{5G}, \qquad u = \frac{m_t + 2m_b}{5G}.
\label{vevs}\end{equation}
In the $\Gamma_D$ mass matrix of the Eq. (\ref{matb}) we made a convenient choice of the signs in order to obtain positive values for the VEVs $v$ and $u$ in Eqs. (\ref{vevs}). We should noticed that this result can also be obtained by an appropriated transformation of the fermion fields {\it via} the $\gamma_5$ matrix. From Eqs. (\ref{vev}) and (\ref{vevs}) we have
\begin{equation}
G = \frac{1}{v_W}\sqrt{\frac{m_t^2 + m_b^2}{5}}.
\label{G}\end{equation}\par
The masses of the other quarks can be induced by one loop radiative corrections scheme. From Eqs. (\ref{mix1}), (\ref{mix2}), (\ref{yukq}) and (\ref{higgs}) we constructed the diagrams of the Figs. 1 and 2. It is easy to see that the contributions of the Fig. 1 to the mass matrix elements of the 2/3 charge sector are
\begin{eqnarray}
m_{ab}^{\left(U\right)} & = & \frac{wG^2}{8\pi^2}{\tt U}_{cb}^R\left\{\frac{v}{v^2 + w^2}\Delta_{1c}{{\tt U}^L_{1b}}^\dagger I\left(m_{J_1}, m_{H_2}\right) + \right.\cr 
&& \left. \frac{u}{u^2 + w^2}\Delta_{\alpha c}\Delta_{\beta\sigma}{{\tt U}^L_{\beta c}}^\dagger\left[{{\tt J}^L_{\alpha2}}^\dagger{\tt J}_{\sigma2}^R\times \right.\right. \cr
&& \left.\left. I\left(m_{J_2}, m_H\right) + {{\tt J}_{\alpha3}^L}^\dagger {\tt J}_{\sigma3}^RI\left(m_{J_3}, m_H\right)\right]\right\},
\label{mu}\end{eqnarray}
\begin{figure}[h]
\begin{center}  
\epsfxsize=2in  
\epsfysize=3 true cm  
\centerline{\epsfbox{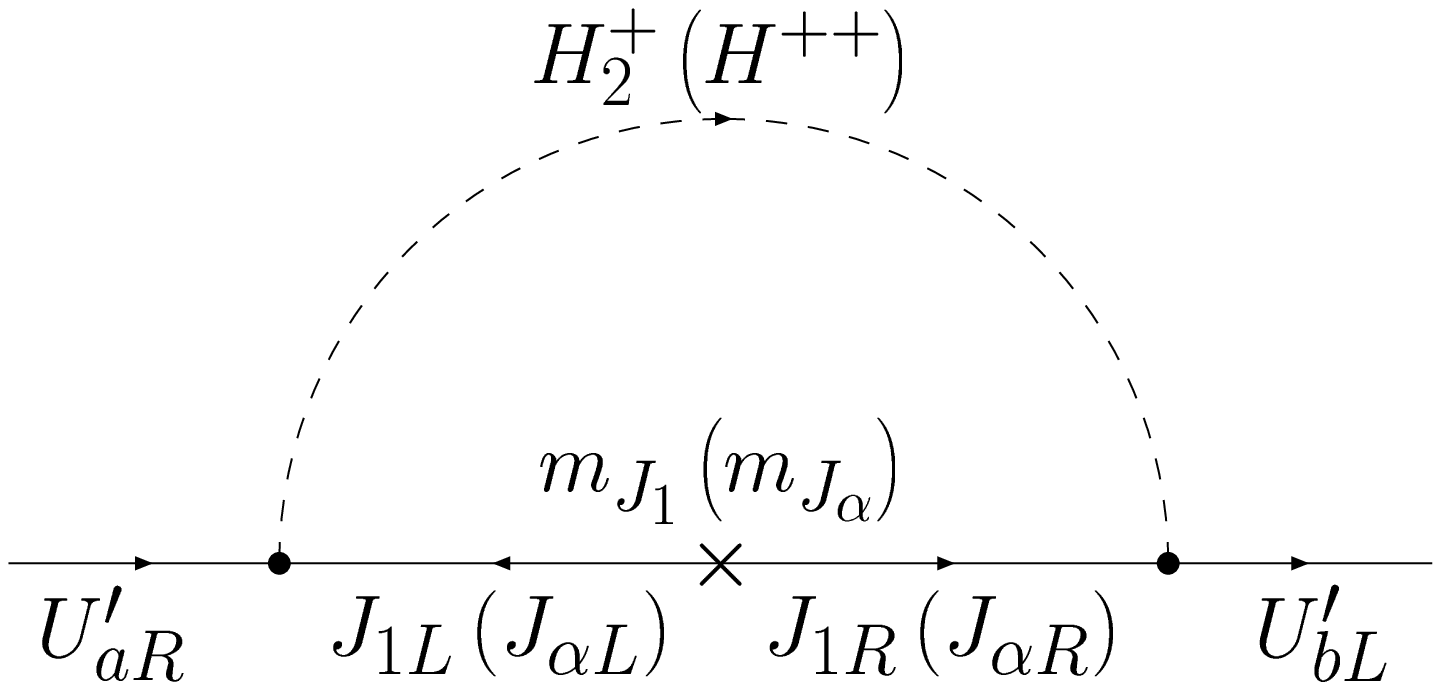}}  
\end{center}  
\caption[]{One loop diagram which contributes to the mass matrix of the quarks of 2/3 charge.}
\label{fig:1a}\end{figure}\noindent
where $m_{H_2}$ and $m_H$ are the masses of one of the single charged and of the double charged Higgs boson, respectively [see Eqs. (\ref{higgs})], $\Delta_{ab} = 1$ and
\begin{equation}
I\left(x, y\right) = \frac{x^3}{x^2 - y^2}\ln{\frac{x}{y}}.
\end{equation}
For the $-$1/3 charge sector the contributions from Fig. 2 are similar, {\it i. e.},
\begin{eqnarray}
m_{ab}^{\left(D\right)} & = & \frac{wG^2}{8\pi^2}{\tt D}_{cb}^R\left\{\frac{u}{u^2 + w^2}\Delta_{1c}{{\tt D}^L_{1a}}^\dagger I\left(m_{J_1}, m_H\right) + \right.\cr 
&& \left. \frac{v}{v^2 + w^2}\Delta_{\alpha c}\Delta_{\beta\sigma}{{\tt D}^L_{\beta c}}^\dagger\left[{{\tt J}^L_{\alpha2}}^\dagger{\tt J}_{\sigma2}^R\times \right.\right. \cr
&& \left.\left. I\left(m_{J_2}, m_{H_2}\right) + {{\tt J}_{\alpha3}^L}^\dagger {\tt J}_{\sigma3}^RI\left(m_{J_3}, m_{H_2}\right)\right]\right\},
\label{md}\end{eqnarray}\par
\begin{figure}[h]
\begin{center}  
\epsfxsize=2in  
\epsfysize=3 true cm  
\centerline{\epsfbox{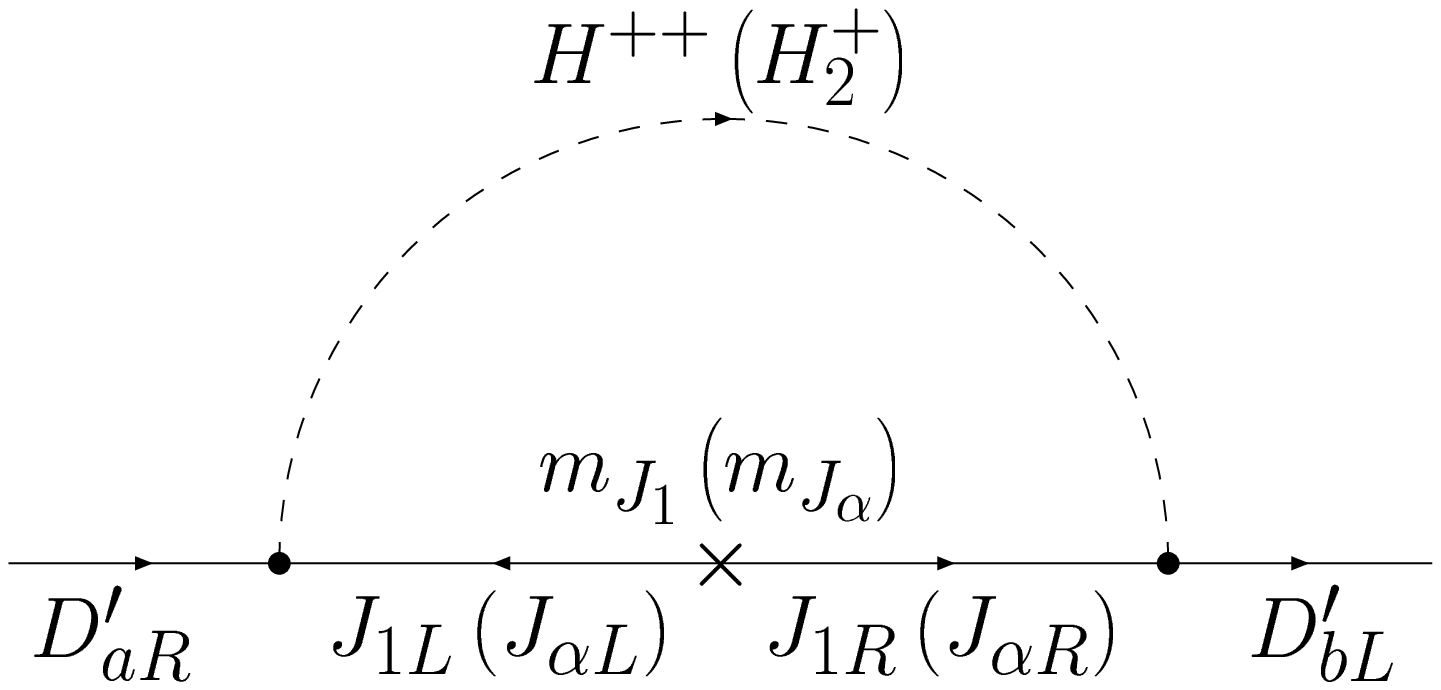}}  
\end{center}  
\caption[]{One loop diagram which contributes to the mass matrix of the quarks of $-$1/3 charge.}
\label{fig:1b}\end{figure}\noindent
Let us discuss now the case of the charged lepton masses. We can see from Eq. (\ref{yukl}) that in the original 3-3-1 heavy lepton model the ordinary charged leptons can get their masses from the $\rho$ Higgs triplet in Eq. (\ref{tri}) at the tree level. However, from Eqs. (\ref{vevs}) the VEV $u$ is $\approx 2v_W/\sqrt{5}$ and from Eq. (\ref{G}) $G \approx 0.3$, with $m_t \sim 174 {\mbox{ GeV}} \gg m_b$ \cite{Gea00}. Therefore, the tree level charged lepton masses would be out of the experimental ranges in our scheme. However, we can introduce a set of discrete symmetries in the leptonic sector, in addition to the ones in Eqs. (\ref{simq}) for the quarks and scalar fields, which forbids tree level charged lepton masses, {\it i. e.},
\begin{mathletters}
\begin{eqnarray}
\psi_{1L} \to i\psi_{1L}, \quad \psi_{\alpha L} & \to & \psi_{\alpha L}, \quad l^\prime_{1R} \to l^\prime_{1R}, \\
l^\prime_{\alpha R} \to -l^\prime_{\alpha R}, \quad P^\prime_{aR} & \to & P^\prime_{aR}.
\end{eqnarray}\label{siml}\end{mathletters}
\begin{figure}[h]
\begin{center}  
\epsfxsize=2in  
\epsfysize=3 true cm  
\centerline{\epsfbox{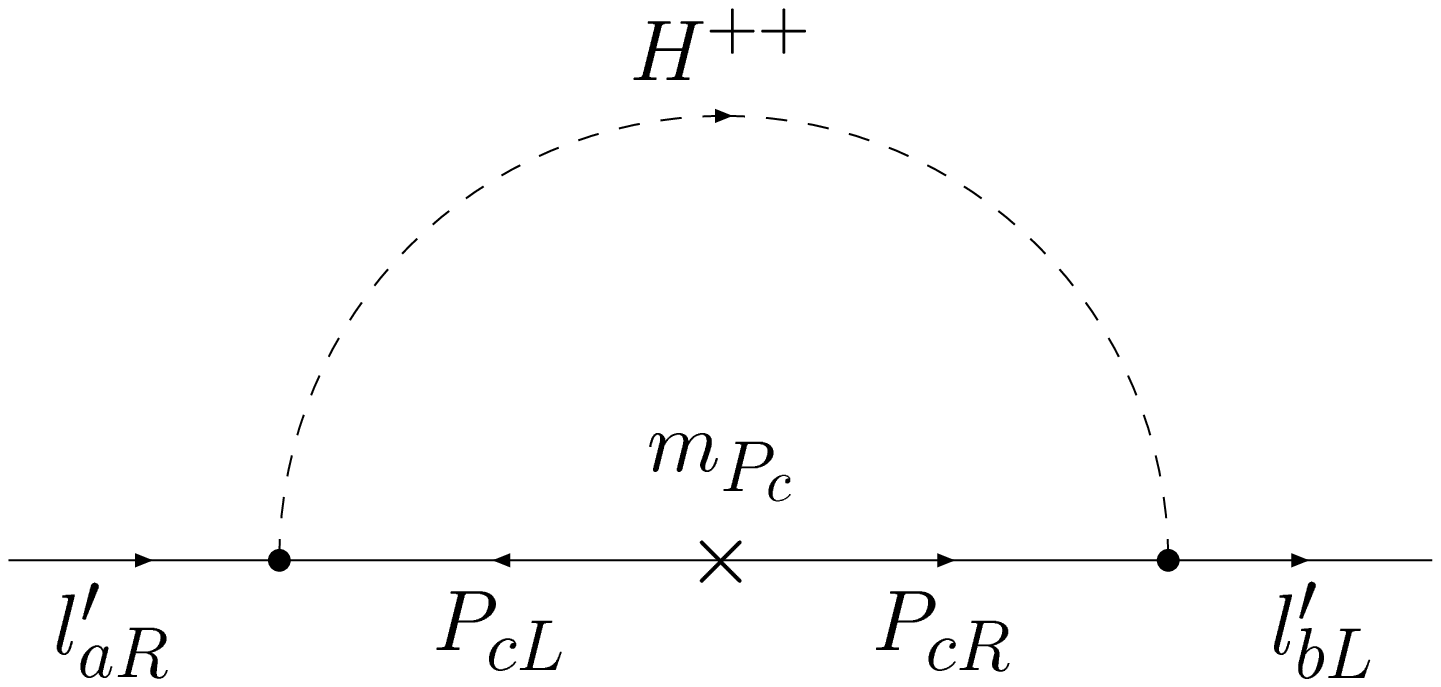}}  
\end{center}  

\caption[]{One loop diagram which contributes to the mass matrix of the charged leptons.}
\label{fig:3}\end{figure}\noindent
In order to implement Yukawa universality in the leptonic sector, we assume also the $S_3$ simmetries within each lepton family and $S_{3L}\otimes S_{3R}$ symmetry among the families. Now, from the Eq. (\ref{yukl}) one realizes that the charged lepton masses at one loop level can be read from Fig. 3. The mass matrix elements for the charged leptons are given by
\begin{eqnarray}
m_{ab}^{\left(l\right)} & = & -\frac{G^2}{8\pi^2}\frac{uw}{u^2 + w^2}\times \cr
&& \sum_cI\left(m_{P_c}, m_H\right)\Delta_{\alpha 1}{{\tt P}^L_{\alpha c}}^\dagger{\tt L}^R_{1b}\Delta_{\beta d}{{\tt L}^L_{\beta a}}^\dagger{\tt P}^R_{dc}.
\label{ml}\end{eqnarray}\noindent
For sake of simplicity we are assuming equal strength of the Yukawa coupling for the quark and the lepton sectors.\par
Let us give a simple numerical example in order to demonstrate the model reliability to reproduce the correct charged fermion mass patterns. For simplicity we assume all the parameters to be real. For the standard model parameters we put $v_W = 246$ GeV, $m_b = 4.3$ GeV and $m_t = 174$ GeV \cite{Gea00}. In addition we assume for the 3-3-1 model $w = 2000$ GeV, $m_{J_a} = Gw$. Considering experimental bounds from other models we impose $m_{H_2} = m_{H} = 50$ GeV as reasonable values \cite{Gea00}. In the charge 2/3 quark sector we take the mixing matrix of the right-handed fields [see Eq. (\ref{mu})] as
\begin{equation}
{\tt U}^R = \pmatrix{0.98 & 0.0010 & 0.0020 \cr 0.0015 & 0.97 & 0.12 \cr 0.00030 & 0.014 & 0.90 \cr}.
\end{equation}
For the left-handed fields we take ${\tt U}_{11}^L = 0.74$, ${\tt U}_{12}^L = 0.020$, ${\tt U}_{13}^L = 0.68$. The other elements of the ${\tt U}^L$ mixing matrix are not necessary in these calculations. In order to obtain the standard model experimental values of the ordinary charged quark masses we assume that the mixing in the second and third terms of the right-handed side in Eq. (\ref{mu}) contribute as
\begin{eqnarray}
{\tt U}^R_{cb}\Delta_{\alpha c}\Delta_{\beta\sigma}{{\tt U}^L_{\beta c}}^\dagger{{\tt J}^L_{\alpha2}}^\dagger{\tt J}_{\sigma2}^R & = & \cr
{\tt U}^R_{cb}\Delta_{\alpha c}\Delta_{\beta\sigma}{{\tt U}^L_{\beta c}}^\dagger{{\tt J}^L_{\alpha3}}^\dagger{\tt J}_{\sigma3}^R & = & 2.64.
\end{eqnarray}
With these values for the parameters we obtain $m_u \approx 3\times 10^{-3}$ GeV and $m_c \approx 1.5$ GeV, respectively, for the up and charm quark masses. The values of the free parameters were choosing in a way that one loop level contribution to the quark top mass vanish. In the $-$1/3 charge sector we take ${\tt D}^R  = {\tt U}^R$, ${\tt D}_{11}^L = -0.34$, ${\tt D}_{12}^L = 0.77$, ${\tt D}_{13}^L = -0.55$ and
\begin{eqnarray}
{\tt D}^R_{cb}\Delta_{\alpha c}\Delta_{\beta\sigma}{{\tt D}^L_{\beta c}}^\dagger{{\tt J}^L_{\alpha2}}^\dagger{\tt J}_{\sigma2}^R & = & \cr
{\tt D}^R_{cb}\Delta_{\alpha c}\Delta_{\beta\sigma}{{\tt D}^L_{\beta c}}^\dagger{{\tt J}^L_{\alpha3}}^\dagger{\tt J}_{\sigma3}^R & = & 0.17.
\end{eqnarray}
These parameters give for the masses of the down and strange quarks $m_d \approx 5 \times 10^{-3}$ GeV and $m_s \approx 0.1$ GeV, respectively. Radiative contribution to the bottom quark mass is zero. \par
Taking the limit in which the mixing term in Eq. (\ref{ml}) is a diagonal matrix we can also reproduce the standard model experimental values for the ordinary charged lepton masses provided the relations $m_{11}^{\left(l\right)} = 0.46m_e$, $m_{22}^{\left(l\right)} = 0.46m_\mu$ and $m_{11}^{\left(l\right)} = 0.46m_\tau$ are valid. For simplicity we considering here equal contribution of the three terms in Eq. (\ref{ml}).\par
Certainly there are several reasonable combination of the free parameters which lead to the correct phenomenology. The numerical example above is one of them which we are employing to illustrate the viability of the model. 

\section{Neutrino masses}

Finally, let us comment on how our scheme for charged fermion masses generation can be connected with a preexisting mechanism for generation of neutrino masses. Models for neutrino masses generation in the 3-3-1 model are discussed in several papers \cite{MP01,OY99}. We show that our scheme can be compatible with the one of the Ref. \cite{OY99}, where the neutrino mass terms can receive three types of contributions. They come from diagrams with ordinary charged leptons, with heavy leptons and from diagrams allowed by mixing between ordinary charged leptons and the heavy ones. If we consider only the contribution of the heavy leptons we have
\begin{equation}
m^{\left(\nu\right)}_{ab} = \lambda \Theta_{ab}\left(m_{P_b}^2 - m_{P_a}^22\right)\frac{vu}{w}I\left(m_{H_1}, m_{H_2}\right),
\label{mnu}\end{equation}
where $m_{H_1}$ and $m_{H_2}$ are masses of the single charged Higgs bosons and $\lambda$ is a parameter of the Higgs potential. Here we are taking lepton mixings into account in the $\Theta_{ab}$ term, which is not present in the model of the Ref. \cite{OY99}. Yukawa coupling constants are also included in this matrix. However, mixing from the scalar sector is not contained in it. Thus, the $\lambda$ constant make the role of the scalar mixing. The VEVs $v$ and $u$ are given by the Eqs. (\ref{vevs}). For guarantee heavy lepton dominance we take $\vert m_{P_b}^2 - m_{P_a}^2\vert = 10$ GeV \cite{OY99}. Assuming Yukawa universality, with coupling constant given by Eq. (\ref{G}) we can write, for simplicity, $\Theta_{ab} = \alpha\Delta_{ab}$. The other numerical parameters are the ones of the charged fermion discussion in Sec. \ref{secIII}. Therefore, from Eq. (\ref{mnu}) we have, for example, $m^{\left(\nu\right)}_{12} \sim m^{\left(\nu\right)}_{13} \sim 0.01\alpha\lambda$ GeV for $m_{P_2} \sim m_{P_3}$. However, to get bi-maximal neutrino mixing we require $m^{\left(\nu\right)}_{12} \sim m^{\left(\nu\right)}_{13} \sim 0.01$ eV and $m^{\left(\nu\right)}_{13}/ m^{\left(\nu\right)}_{23} \sim \Delta m^2_{\rm atm}/\Delta m^2_{\rm sol}$, with $\Delta m^2_{\rm atm} \sim 10^{-3}$ eV$^2$ and $\Delta m^2_{\rm sol} \sim 10^{-5}$ eV$^2$ or $10^{-10}$ eV$^2$ \cite{Gea00}. Therefore, from Eq. (\ref{mnu}) we must have $\alpha\lambda \sim 10^{-7}$ and $m_{23}^{\left(\nu\right)} \sim 10^{5}\alpha\lambda \sim 10^{-2}$ eV. These values for the neutrino masses are in agreement with the recent data on solar and atmospheric neutrinos oscillations \cite{Aea99}.

\section{Conclusion}

In conclusion, we implemented a model for generation of ordinary charged fermion masses in the context of the 3-3-1 heavy lepton model \cite{PT93}. The scheme is able to reproduces the observed hierarchy of the ordinary charged fermion masses without hierarchy in the Yukawa coupling strength. In the model the spontaneous symmetry breaking occurs in two steps, SU(3)$\otimes$U(1) $\to$ SU(2)$\otimes$U(1), governed by the VEV $w$, and SU(2)$\otimes$U(1) $\to$ U(1), governed by $v$ and $u$. The VEV $w$ is responsible for the masses of the exotic fermions and $v$ and $u$ for the ordinary charged fermion masses. It is required in this work that $v$ and $u$ are of the same order of magnitude [see Eqs. (\ref{vevs})]. Only the top and the bottom quarks get masses at the tree level. The difference among the masses in each charge sector is controlled by the mixing parameters. It is possible to show, by a reasonable choice of the free parameters, that some masses can be very smaller that the electroweak scale. Our scheme employs $S_3$ permutation simmetries and the other sets of discrete symmetries in Eqs. (\ref{simq}) and (\ref{siml}). We remember that permutation symmetry was already employed for studies of the fermion mass and mixing problems in context of other models \cite{BM90}. In our mechanism, compared with the standard model and other extended models, it is easier to understand why the top and the bottom quarks are the heavier known fermions and why the top quark mass is the only one near to the Fermi scale. We show also that the scheme is compatible with a mechanism for neutrino mass generation at one loop level proposed by other authors \cite{OY99}. 

\acknowledgements
I would like to thank the Instituto de F\'\i sica Te\'orica, UNESP, for the use of its facilities, the Funda\c c\~ao de Amparo \`a Pesquisa no Estado de S\~ao Paulo (Processo No. 99/07956-3), for full financial support and the Dr. M. M. Leite for a critical reading of the manuscript.

\end{document}